\documentclass[5p]{elsarticle}

\usepackage{graphicx}
\usepackage{amsmath}
\usepackage{amssymb}

\usepackage{bm}
\journal{Journal of Magnetism and Magnetic Materials}

\DeclareMathOperator{\sgn}{sgn}

\DeclareMathOperator{\Sp}{\mathrm{Sp}}
\begin{document}

\begin{frontmatter}

\title{Giant thermopower in superconducting heterostructures with spin-active
interfaces}

\author[LPI,LCN]{Mikhail S. Kalenkov}
\author[KIT,LPI]{Andrei D. Zaikin}

\address[LPI]{I.E. Tamm Department of Theoretical Physics, P.N. Lebedev Physical Institute, 119991 Moscow, Russia}
\address[KIT]{Institut f\"ur Nanotechnologie, Karlsruher Institut f\"ur Technologie
(KIT), 76021 Karlsruhe, Germany}
\address[LCN]{Laboratory of Cryogenic Nanoelectronics, Nizhny Novgorod State Technical University, 603950 Nizhny Novgorod, Russia}

\begin{abstract}
We predict parametrically strong enhancement of the thermoelectric effect
in metallic bilayers consisting of two superconductors separated by a spin-active interface. The physical mechanism for such an enhancement is
directly related to electron-hole imbalance generated by spin-sensitive quasiparticle scattering at the interface between superconducting layers. 
We explicitly evaluate the thermoelectric currents flowing in the system
and demonstrate that they can reach maximum values comparable to the critical ones for superconductors under consideration.
\end{abstract}

\begin{keyword}
Superconductivity, thermoelectric effect, spin-dependent electron scattering,
electron-hole imbalance
\end{keyword}

\end{frontmatter}


\section{Introduction}
It is well known that application of a thermal gradient $\nabla T$ to a normal
conductor along with electric field $\bm{E}$ results in the electric current
\begin{equation}
\bm{j}=\sigma_N \bm{E} + \alpha_N \nabla T, \quad  \alpha_N \sim 
(\sigma_N/e)(T/\varepsilon_F).
\label{norm_resp}
\end{equation}
Here $\sigma_N$ defines Drude conductivity, $\alpha_N$ is
thermoelectric coefficient and $\varepsilon_F$ is the Fermi energy.
Provided a metal is brought into a superconducting state, Eq. \eqref{norm_resp}
is no longer correct since the electric field cannot penetrate into the bulk of
a superconductor. Instead, one finds
\begin{equation}
\bm{j}=\bm{j}_s + \alpha_S \nabla T,
\label{super_resp}
\end{equation}
where $\bm{j}_s$ is a supercurrent and $\alpha_S$ defines thermoelectric
coefficient in a superconducting state. It turns out that by applying
thermal gradient to a uniform superconductor it is not possible to induce
and measure any current since thermal current would always be compensated by the
supercurrent $\bm{j}_s =- \alpha_S \nabla T$. The way out is to
consider non-uniform superconducting structures in which case no such compensation generally
occurs \cite{Ginzburg44,Ginzburg91} and the thermoelectric current can be detected
experimentally. Making use of this idea the thermoelectric effect was indeed demonstrated
in several experiments with bimetallic superconducting rings
\cite{Zavaritskii74,Falco76,Harlingen80}. Quite surprisingly, the magnitude of the effect was found to be 
several orders of magnitude bigger than predicted by theory \cite{Galperin73}. 
The authors of a very recent experimental work \cite{pe} also observed a discrepancy between theory 
and their experimental data.

By now it is well understood that a small theoretical value of the thermoelectric coefficient
in ordinary superconductors \cite{Galperin73} $\alpha_S \sim \alpha_N$  is directly linked to
the assumption that electron-hole symmetry remains preserved in these structures.
In this case contributions to the thermoelectric
current provided by electron-like and hole-like excitations are of
the opposite sign and almost cancel each other. Then, like in a normal metal,
one inevitably finds that $\alpha_S$ is controlled by a parametrically small
factor $T/\varepsilon_F \ll 1$.

The situation may change if for some reason
the electron-hole symmetry gets violated. In this case -- as it was demonstrated by
a number of authors -- a much stronger thermoelectric effect can be expected.
The proposed mechanisms for the electron-hole symmetry violation and the related
thermoelectric effect enhancement are diverse. In conventional superconductors doped by magnetic impurities, the presence of Andreev
bound states formed near such impurities may yield
an asymmetry between electron and hole scattering rates which in turn
results in a drastic enhancement of the thermoelectric effect \cite{Kalenkov12}.
Likewise, the formation of quasi-bound Andreev states near
non-magnetic impurities in unconventional superconductors may lead to much larger
values of $\alpha_S$ in such systems \cite{LF}. Substantial enhancement of thermoelectric currents was also
predicted in three terminal hybrid ferromagnet-superconductor-ferromagnet (FSF) \cite{Machon13}
as well as in FS junctions in the presence of a Zeeman spin-splitting field \cite{Ozaeta13}.

In a recent work \cite{Kalenkov14} we argued that the thermoelectric effect can be strongly enhanced
also in metallic bilayers consisting of a superconductor and a normal metal (SN) provided these two
metals are separated by a thin spin-active interface. By exactly solving the corresponding Bogolyubov-de-Gennes
equations we evaluated the wave functions for electron-like and hole-like excitations in such systems
demonstrating that spin-sensitive scattering at the SN interface can generate electron-hole imbalance
and result in the presence of large thermoelectric currents in such systems.
In this paper we will further extend our arguments \cite{Kalenkov14} to superconducting multilayers with spin-active interfaces
and demonstrate that thermoelectric properties of such systems may drastically differ from the those of bulk superconductors.
As a simple example of such systems below we will specifically consider a superconductor with a thin
ferromagnetic interlayer. We will show that provided a temperature gradient is applied along this interlayer
the system develops a thermoelectric current which maximum values can be as high as the critical (depairing)
current of a superconductor.

The structure of our paper is as follows. In Sec. \ref{formalism} we will specify our model and outline our basic quasiclassical formalism
of Eilenberger equations to be employed in our analysis of the thermoelectric effect. In Sec. \ref{Riccati} we will present
an efficient method enabling one to derive the solution of these equations for the system under consideration. With the aid of this solution
we will then derive a general expression for the thermoelectric current and also briefly discuss our results in Sec. \ref{thermocurrent}.

\begin{figure}
\centerline{ \includegraphics[width=80mm]{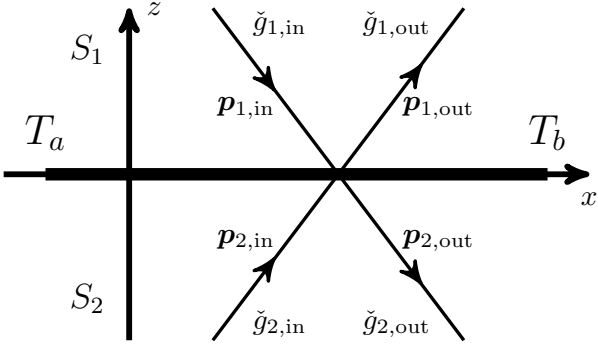} }
\caption{The system under consideration consisting of two superconducting layers 
$S_1$ and $S_2$ separated by a spin-active interface. The left and right edges 
of this superconducting bilayer are maintained at temperatures $T_a$ and $T_b$, 
respectively. We also schematically indicate the quasiclassical electron Green functions for incoming and 
outgoing momentum values. These Green functions are matched at the spin-active interface by means of the 
proper boundary conditions as specified in the text.}
\label{sis-fig}
\end{figure}

\section{The model and quasiclassical formalism}
\label{formalism}
In what follows we will consider an extended metallic bilayer consisting of the two superconducting slabs $S_1$ and $S_2$ as shown in Fig. \ref{sis-fig}. We
will assume that both metals are brought into direct contact with each other via a spin-active interface that is located in the plane $z=0$. Such an interface can be formed, e.g., by an ultrathin layer of a ferromagnet.
Our goal is to evaluate an electric current response to a temperature gradient applied to the system along the $S_1 S_2$ interface. This temperature gradient
is achieved by setting the temperature $T$ at the left ($x \to -\infty$) and right ($x \to \infty$) edges of the bilayer equal respectively to $T_a$ and $T_b$, see Fig. \ref{sis-fig}.
For the sake of simplicity below we will assume that the temperature depends only on $x$ and does not vary along $y$- and $z$-directions.

Within the quasiclassical theory of superconductivity \cite{bel}, the current density $\bm{j}(\bm{r})$ in our system can be
evaluated by means of the standard formula
\begin{equation}
\bm{j}(\bm{r})= -\dfrac{e N_0}{8} \int d \varepsilon
\left< \bm{v}_F \Sp [\hat \tau_3 \hat g^K(\bm{p}_F, \bm{r},
\varepsilon) ] \right>,
\label{current}
\end{equation}
where $N_0$ is the density of state at the Fermi level, $\bm{p}_F=m\bm{v}_F$ is the electron Fermi momentum vector,
$\hat\tau_3$ is the Pauli matrix in the Nambu space,
the angular brackets $\left<\cdots\right>$ denote averaging over the Fermi momentum directions and $\hat g^K$ is the Keldysh block of the
quasiclassical Green-Eilenberger function matrix
\begin{equation}
\check g =
\begin{pmatrix}
\hat g^R & \hat g^K \\
0 & \hat g^A
\end{pmatrix}.
\label{green}
\end{equation}
Here and below the ``hat''-symbol denotes $4\times4$ matrices in the 
Nambu$\otimes$Spin space while the ``check''-symbol
labels $8\times8$ matrices in the Keldysh$\otimes$Nambu$\otimes$Spin space.

The matrix function $\check g$ obeys the transport-like Eilenberger equation \cite{bel}
\begin{equation}
\left[ \varepsilon \hat\tau_3 - \check\Delta(\bm{r}), \check g \right]
+
i\bm{v}_F \nabla \check g (\bm{p}_F, \bm{r}, \varepsilon) =0
\label{Eil2}
\end{equation}
as well as the normalization condition
\begin{equation}
\check g^2 =1.
\label{norm}
\end{equation}
The order parameter matrix $\check \Delta$ has only ``retarded'' and ``advanced'' components,
\begin{equation}
\check \Delta =
\begin{pmatrix}
\hat \Delta & 0\\
0 & \hat \Delta
\end{pmatrix}, \quad \hat \Delta
=
\begin{pmatrix}
0 & \Delta \sigma_0 \\
-\Delta^* \sigma_0 & 0
\end{pmatrix},
\label{chde}
\end{equation}
where $\sigma_0$ is the unity matrix in the spin space and $\Delta$ is the superconducting order parameter. As soon as we are interested in the electronic transport along the interface we set the phase difference between the two superconductors $S_1$ and $S_2$ to zero. Under this assumption order parameter can be made to be real everywhere in the system.

As usually, the quasiclassical equations \eqref{Eil2} should be supplemented
by boundary conditions
which describe electron transfer across the $SFS$-interface by matching the Green function matrices $\check g$ for incoming and outgoing momentum directions
at both sides of this interface, see Fig. \ref{sis-fig}. In the case of
spin-active interfaces the corresponding boundary conditions
were derived in \cite{Millis88}. Here we will employ an equivalent approach \cite{Zhao04}.

The simplest model of the spin-active interface is described by three
parameters, i.e. the transmission probabilities for opposite spin directions
$D_{\uparrow}$ and $D_{\downarrow}$ as well as the so-called
spin mixing angle $\theta$. Previously we have already made use of this model, e.g., 
while considering crossed Andreev reflection in three-terminal FSF structures \cite{KZ07} or 
triplet pairing and dc Josephson effect in SFS junctions \cite{KGZ09}. For simplicity we assume 
that the above three parameters
do not depend on the sign of the quasiparticle momentum along the interface,
i.e. $D_{\uparrow}(\bm{p}_{\parallel})=D_{\uparrow}(-\bm{p}_{\parallel})$ and so on. Then the
elements of the interface scattering matrices for electrons
\begin{equation}
\mathcal{S}
=
\begin{pmatrix}
S_{11} & S_{12} \\
S_{21} & S_{22}
\end{pmatrix}
\end{equation}
and holes
\begin{equation}
\underline{\mathcal{S}}
=
\begin{pmatrix}
\underline{S}_{11} & \underline{S}_{12} \\
\underline{S}_{21} & \underline{S}_{22}
\end{pmatrix}
\end{equation}
take the form
\begin{gather}
S_{11}=S_{22}=
\sqrt{R_{\sigma}}e^{i\theta_{\sigma}/2},
\\
S_{12}=S_{21}=
i \sqrt{D_{\sigma}}e^{i\theta_{\sigma}/2},
\\
\underline{S}_{11}=\underline{S}_{22}=
\sqrt{R_{-\sigma}}e^{-i\theta_{\sigma}/2},
\\
\underline{S}_{12}=\underline{S}_{21}=
i \sqrt{D_{-\sigma}}e^{-i\theta_{\sigma}/2},
\end{gather}
where $\theta_{\sigma}$ is $2\times 2$ diagonal matrix in the spin space defined as $\theta_{\sigma}=\theta \sigma_3$.
The matrices $D_{\pm\sigma}$ and $R_{\pm\sigma}$ are composed of transmission and reflection probabilities for opposite spin directions as
\begin{gather}
D_{\sigma}=
\begin{pmatrix}
D_{\uparrow} & 0 \\
0 & D_{\downarrow}
\end{pmatrix},
\quad
D_{-\sigma}=
\begin{pmatrix}
D_{\downarrow} & 0 \\
0 & D_{\uparrow}
\end{pmatrix},
\\
R_{\sigma}=
\begin{pmatrix}
R_{\uparrow} & 0 \\
0 & R_{\downarrow}
\end{pmatrix},
\quad
R_{-\sigma}=
\begin{pmatrix}
R_{\downarrow} & 0 \\
0 & R_{\uparrow}
\end{pmatrix},
\end{gather}
where we defined the spin-up and spin-down reflection coefficients respectively as 
$R_{\uparrow}=1-D_{\uparrow}$ and $R_{\downarrow}=1-D_{\downarrow}$.

\section{Riccati parameterization}
\label{Riccati}

In order to proceed we will employ the so-called Riccati parameterization of the retarded
and advanced Green functions\cite{Schopohl95,Eschrig00}.
\begin{equation}
\hat g^{R,A}=\pm
    \hat N^{R,A}
    \begin{pmatrix}
    1+\gamma^{R,A} \tilde \gamma^{R,A} & 2\gamma^{R,A} \\
    -2 \tilde \gamma^{R,A} & -1- \tilde \gamma^{R,A}  \gamma^{R,A} \\
    \end{pmatrix},
    \label{graparam}
\end{equation}
where $\hat N^{R,A}$ represent the following matrices
\begin{equation}
\hat N^{R,A}=
    \begin{pmatrix}
    (1-\gamma^{R,A} \tilde \gamma^{R,A})^{-1} & 0 \\
    0 & (1-\tilde \gamma^{R,A}  \gamma^{R,A} )^{-1} \\
    \end{pmatrix}.
    \label{nrparam}
\end{equation}
Here the Riccati amplitudes $\gamma^{R,A}$, $\tilde \gamma^{R,A}$ are $2\times2$ matrices in the spin space.

Parameterization of the Keldysh Green function involves two distribution functions \cite{Eschrig00} $x^K$, $\tilde x^K$ also being $2\times2$ matrices in the spin space, namely
\begin{equation}
\hat g^K=
2
\hat N^R
\begin{pmatrix}
x^K - \gamma^R  \tilde x^K  \tilde \gamma^A &
-\gamma^R  \tilde x^K + x^K  \gamma^A \\
-\tilde \gamma^R  x^K + \tilde x^K  \tilde \gamma^A &
\tilde x^K - \tilde \gamma^R  x^K  \gamma^A \\
\end{pmatrix}
\hat N^A.
\label{gkparam}
\end{equation}

The amplitudes $\gamma^{R,A}$, $\tilde \gamma^{R,A}$ obey the Riccati equations
\begin{gather}
i\bm{v}_F \nabla \gamma^{R,A}=
\begin{pmatrix}
1 & \gamma^{R,A}
\end{pmatrix}
\hat h
\begin{pmatrix}
-\gamma^{R,A} \\ 1
\end{pmatrix},
\label{ric}
\\
i\bm{v}_F \nabla \tilde \gamma^{R,A}=
\begin{pmatrix}
\tilde \gamma^{R,A}& 1
\end{pmatrix}
\hat h
\begin{pmatrix}
1 \\  -\tilde \gamma^{R,A}
\end{pmatrix},
\label{tilderic}
\end{gather}
while the distribution functions $x^K$ and $\tilde x^K$ satisfy the transport-like equations
\begin{multline}
i\bm{v}_F \nabla x^K
=
x^K
\begin{pmatrix}
1 & 0 \\
\end{pmatrix}
\hat h
\begin{pmatrix}
1 \\ -\tilde\gamma^A \\
\end{pmatrix}
-
\begin{pmatrix}
1 & \gamma^R \\
\end{pmatrix}
\hat h
\begin{pmatrix}
1 \\ 0 \\
\end{pmatrix}
x^K,
\end{multline}
and
\begin{multline}
i\bm{v}_F \nabla \tilde x^K
=
\tilde x^K
\begin{pmatrix}
0 & 1 \\
\end{pmatrix}
\hat h
\begin{pmatrix}
-\gamma^A \\ 1 \\
\end{pmatrix}
-
\begin{pmatrix}
\tilde \gamma^R & 1 \\
\end{pmatrix}
\hat h
\begin{pmatrix}
0 \\ 1 \\
\end{pmatrix}
\tilde x^K,
\end{multline}
where $\hat h= \varepsilon \hat\tau_3 - \hat \Delta(\bm{r})$.

Below it will be convenient for us to employ the parameterization of the Green function matrices
in terms of the Riccati amplitudes $\gamma$, $\tilde \gamma$, $\Gamma$, $\tilde \Gamma$ as well as the
distribution functions $x$, $\tilde x$, $X$, $\tilde X$ (all being $2\times 2$ matrices in the spin space) \cite{Eschrig00},
\begin{gather}
\check g_{i, \text{in}} = \check g_{i, \text{in}}
[\gamma_i^R, \tilde \Gamma_i^R, \Gamma_i^A, \tilde \gamma_i^A, x_i, \tilde X_i],
\quad i=1,2,
\\
\check g_{i, \text{out}} = \check g_{i, \text{out}}
[\Gamma_i^R, \tilde \gamma_i^R, \gamma_i^A, \tilde \Gamma_i^A, X_i, \tilde x_i],
\quad i=1,2.
\end{gather}

Boundary conditions\cite{Zhao04} allow to express the interface values of the capital
functions $\Gamma$ and $X$ in terms of the lower-case functions $\gamma$ and $x$. For Riccati amplitudes $\Gamma^{R,A}_{1}$ and $\tilde \Gamma^{R,A}_{1}$ at the interface we obtain
\begin{multline}
\Gamma_1^R(0)=
\tilde \Gamma_1^R(0)=
\mathcal{M}_1^R
e^{i\theta_{\sigma}}
\Bigl\{
\gamma^R_1(0)\sqrt{R_{\uparrow} R_{\downarrow}}
\\+
\gamma^R_2(0)\sqrt{D_{\uparrow} D_{\downarrow}}
-
\gamma^R_1(0) \left[\gamma^R_2(0)\right]^2 e^{i\theta_{\sigma}}
\Bigr\},
\label{ricR1int}
\end{multline}
\begin{multline}
\Gamma_1^A(0)=
\tilde \Gamma_1^A(0)=
\mathcal{M}_1^A
e^{-i\theta_{\sigma}}
\Bigl\{
\gamma^A_1(0)\sqrt{R_{\uparrow} R_{\downarrow}}
\\+
\gamma^A_2(0)\sqrt{D_{\uparrow} D_{\downarrow}}
-
\gamma^A_1(0) \left[\gamma^A_2(0)\right]^2 e^{-i\theta_{\sigma}}
\Bigr\},
\label{ricA1int}
\end{multline}
where
\begin{multline}
\mathcal{M}_1^R=
\Bigl\{
1-
\left[\gamma^R_2(0)\right]^2 \sqrt{R_{\uparrow} R_{\downarrow}} e^{i\theta_{\sigma}}
\\-
\gamma^R_2(0) \gamma^R_1(0) \sqrt{D_{\uparrow} D_{\downarrow}} e^{i\theta_{\sigma}}
\Bigr\}^{-1},
\label{MR1}
\end{multline}
and
\begin{multline}
\mathcal{M}_1^A=
\Bigl\{
1-
\left[\gamma^A_2(0)\right]^2 \sqrt{R_{\uparrow} R_{\downarrow}} e^{-i\theta_{\sigma}}
\\-
\gamma^A_2(0) \gamma^A_1(0) \sqrt{D_{\uparrow} D_{\downarrow}} e^{-i\theta_{\sigma}}
\Bigr\}^{-1}.
\label{MA1}
\end{multline}
The interface values of the Riccati amplitudes $\Gamma^{R,A}_{2}$ and $\tilde \Gamma^{R,A}_{2}$ can be obtained from Eqs. \eqref{ricR1int}-\eqref{MA1} by interchanging the indices $1\leftrightarrow 2$. Here we used the fact that for the real order parameter one has $\gamma_i^{R,A} = \tilde \gamma_i^{R,A}$. From Eqs. \eqref{ricR1int}-\eqref{MA1} we observe that within the adopted model for the spin-active interface this equality also holds for the capital Riccati amplitudes $\Gamma_i^{R,A} = \tilde \Gamma_i^{R,A}$

In the same way we can express the interface values of the distribution functions. At the $S_1$-side of the interface we have
\begin{multline}
X_1(0)=
\mathcal{M}_1^R \mathcal{M}_1^A
\Biggl(
x_1^K(0)
\left\{ \sqrt{R_{\sigma}} - \left[\gamma^R_2(0)\right]^2 \sqrt{R_{-\sigma}} e^{i\theta_{\sigma}}
\right\}
\\\times
\left\{ \sqrt{R_{\sigma}} - \left[\gamma^A_2(0)\right]^2 \sqrt{R_{-\sigma}} e^{-i\theta_{\sigma}} \right\}
\\+
x_2^K(0)
\left\{ \sqrt{D_{\sigma}} - \gamma^R_1(0) \gamma^R_2(0) \sqrt{D_{-\sigma}} e^{i\theta_{\sigma}}
\right\}
\\\times
\left\{ \sqrt{D_{\sigma}} - \gamma^A_1(0) \gamma^A_2(0) \sqrt{D_{-\sigma}} e^{-i\theta_{\sigma}} \right\}
\\-
\tilde x_2^K(0)
\left\{ \gamma^R_1(0) \sqrt{R_{\sigma}D_{-\sigma}} -
\gamma^R_2(0) \sqrt{R_{-\sigma}D_{\sigma}} \right\}
\\\times
\left\{ \gamma^A_1(0) \sqrt{R_{\sigma}D_{-\sigma}} -
\gamma^A_2(0) \sqrt{R_{-\sigma}D_{\sigma}} \right\}
\Biggr),
\label{X1}
\end{multline}
and
\begin{multline}
\tilde X_1(0)=
\mathcal{M}_1^R \mathcal{M}_1^A
\Biggl(
\tilde x_1^K(0)
\left\{ \sqrt{R_{-\sigma}} - \left[\gamma^R_2(0)\right]^2 \sqrt{R_{\sigma}} e^{i\theta_{\sigma}}
\right\}
\\\times
\left\{ \sqrt{R_{-\sigma}} - \left[\gamma^A_2(0)\right]^2 \sqrt{R_{\sigma}} e^{-i\theta_{\sigma}} \right\}
\\+
\tilde x_2^K(0)
\left\{ \sqrt{D_{-\sigma}} - \gamma^R_1(0) \gamma^R_2(0) \sqrt{D_{\sigma}} e^{i\theta_{\sigma}}
\right\}
\\\times
\left\{ \sqrt{D_{-\sigma}} - \gamma^A_1(0) \gamma^A_2(0) \sqrt{D_{\sigma}} e^{-i\theta_{\sigma}} \right\}
\\-
x_2^K(0)
\left\{ \gamma^R_1(0) \sqrt{R_{-\sigma}D_{\sigma}} -
\gamma^R_2(0) \sqrt{R_{\sigma}D_{-\sigma}} \right\}
\\\times
\left\{ \gamma^A_1(0) \sqrt{R_{-\sigma}D_{\sigma}} -
\gamma^A_2(0) \sqrt{R_{\sigma}D_{-\sigma}} \right\}
\Biggr).
\label{tildeX1}
\end{multline}
The interface values for the distribution functions $X_2(0)$ and $\tilde X_2(0)$ can be recovered from Eqs. \eqref{X1} and \eqref{tildeX1} simply by interchanging the indices $1\leftrightarrow 2$.
Within our simple model all Riccati amplitudes and distribution functions are diagonal in the spin space. The coordinate dependence of the distribution function can be easily found. We obtain
\begin{gather}
x_i = \left[1- \gamma_i^R \gamma_i^A\right]
\times
\begin{cases}
h_a, & v_{x} > 0,
\\
h_b, & v_{x} < 0,
\end{cases}
\label{coord1}
\\
\tilde x_i = - \left[1- \gamma_i^R \gamma_i^A\right]
\times
\begin{cases}
h_b, & v_{x} > 0,
\\
h_a, & v_{x} < 0,
\end{cases},
\label{coord2}
\\
X_i = X_i(0) \dfrac{1-\Gamma_i^R\Gamma_i^A}{1-\Gamma_i^R(0)\Gamma_i^A(0)},
\label{coord3}
\\
\tilde X_i = \tilde X_i(0)\dfrac{1-\Gamma_i^R\Gamma_i^A}{1-\Gamma_i^R(0)\Gamma_i^A(0)},
\label{coord4}
\end{gather}
where the functions $h_{a,b}$ are related to the equilibrium (Fermi) distribution function with temperatures $T_{a,b}$, i.e.
\begin{equation}
h_{a,b}=\tanh \dfrac{\varepsilon}{2T_{a,b}}.
\end{equation}

\section{Thermoelectric current}
\label{thermocurrent}

Let us apply the above quasiclassical formalism in order to derive the thermoelectric current flowing along the spin-active interface in the ballistic limit.
According to Eq. \eqref{current} the component of the current density along
the interface is expressed in terms of the combination $\Sp(\hat \tau_3 \hat
g^K_{\text{in}} + \hat \tau_3 \hat g^K_{\text{out}})$. The above combination is
evaluated by solving the Eilenberger equations \eqref{Eil2}, \eqref{norm}
supplemented by the proper boundary conditions at the $S_1S_2$ interface.
This task can be conveniently accomplished employing the
Riccati parameterization of the Green functions \cite{Schopohl95,Eschrig00}.
Making use of the results of the previous section we obtain
\begin{multline}
\Sp(\hat \tau_3 \hat g^K_{i, \text{in}} + \hat \tau_3 \hat g^K_{i, \text{out}})
=2\Sp\left[
\dfrac{(x_i-\tilde x_i) (1+ \Gamma^R_i \Gamma^A_i)}{
\left(1- \gamma^R_i \Gamma^R_i\right)\left(1- \gamma^A_i
\Gamma^A_i\right)}\right]
\\+
2\Sp\left[
\dfrac{(X_i- \tilde X_i) (1+\gamma^R_i \gamma^A_i)}{
\left(1- \gamma^R_i \Gamma^R_i\right)
\left(1- \gamma^A_i
\Gamma^A_i\right)}\right], \quad i=1,2.
\end{multline}
With the aid of Eqs. \eqref{coord1}-\eqref{coord3} one can rewrite the above
expression in the form
\begin{multline}
\Sp(\hat \tau_3 \hat g^K_{i, \text{in}} + \hat \tau_3 \hat g^K_{i, \text{out}})
\\=2\Sp\left[
\dfrac{\left(1-\gamma^R_i\gamma^A_i\right) (1+ \Gamma^R_i \Gamma^A_i)}{
\left(1- \gamma^R_i \Gamma^R_i\right)\left(1- \gamma^A_i \Gamma^A_i\right)}
\right](h_a + h_b)
\\+2\Sp\left[
\dfrac{ (1-\Gamma^R_i\Gamma^A_i )(1+\gamma^R_i \gamma^A_i)}{
\left(1- \gamma^R_i \Gamma^R_i\right)\left(1- \gamma^A_i \Gamma^A_i\right)}
\dfrac{X_i(0) - \tilde X_i(0)}{1-\Gamma^R_i(0)\Gamma^A_i(0)}
\right].
\label{eq1}
\end{multline}
Eq. \eqref{eq1} can further be simplified if one observes that the four
functions $\gamma^R_i$, $1/\gamma^A_i$, $1/\Gamma^R_i$ and $\Gamma^A_i$ obey
the same Riccati equations implying that the combination
$$
\dfrac{1-\Gamma^R_i \Gamma^A_i}{1-\Gamma^R_i \gamma^R_i}
\dfrac{1-\gamma^A_i\gamma^R_i}{1-\gamma^A_i\Gamma^A_i}
$$
is spatially constant, i.e. it does not depend on the coordinates. Then one can rewrite Eq. \eqref{eq1} as
\begin{multline}
\Sp(\hat \tau_3 \hat g^K_{i, \text{in}} + \hat \tau_3 \hat g^K_{i, \text{out}})
=
2\Sp\left[
\dfrac{\left(1-\gamma^R_i\gamma^A_i\right) (1+ \Gamma^R_i \Gamma^A_i)}{
\left(1- \gamma^R_i \Gamma^R_i\right)\left(1- \gamma^A_i \Gamma^A_i\right)}
\right]
\\\times
(h_a + h_b)
+
2\dfrac{1+\gamma^R_i \gamma^A_i}{1-\gamma^R_i\gamma^A_i}
\\\times
\Sp\left\{
\dfrac{ \left[ 1-\gamma^A_i(0) \gamma^R_i(0)\right]
\left[X_i(0) - \tilde X_i(0)\right]
}{
\left[1- \gamma^R_i(0) \Gamma^R_i(0)\right]\left[1- \gamma^A_i(0)
\Gamma^A_i(0)\right]}
\right\}.
\label{speq}
\end{multline}
What remains is to find the difference of the distribution functions $X_i(0) -
\tilde X_i(0)$ on both sides of the spin-active interface. Making use of Eqs.
\eqref{X1}-\eqref{tildeX1}, we obtain
\begin{multline}
X_i(0)- \tilde X_i(0)=
\mathcal{M}_i^R \mathcal{M}_i^A
\\\times
\left[
A_i(h_a-h_b)\sigma_3 \sgn v_{x} +B_i(h_a+h_b)\right], \quad i=1,2,
\label{AB}
\end{multline}
where
\begin{multline}
A_1=\left(R_{\downarrow} - R_{\uparrow} \right)
K_2
\left[\gamma_1^R(0) \gamma_1^A(0) - \gamma_2^R(0) \gamma_2^A(0)\right],
\label{A1}
\end{multline}
\begin{multline}
B_1=K_2
\left[1+ \gamma_1^R(0) \gamma_1^A(0) \gamma_2^R(0) \gamma_2^A(0)\right]
\\+
\left(R_{\uparrow} + R_{\downarrow} \right)
K_1\gamma_2^R(0) \gamma_2^A(0)
\\-
\sqrt{R_{\uparrow} R_{\downarrow}}
K_1\left\{
\left[\gamma_2^R(0)\right]^2 e^{i\theta_{\sigma}}
+
\left[\gamma_2^A(0)\right]^2 e^{-i\theta_{\sigma}}
\right\}
\\-
\sqrt{D_{\uparrow} D_{\downarrow}}
K_2
\left\{
\gamma_1^R(0)\gamma_2^R(0) e^{i\theta_{\sigma}}
+
\gamma_1^A(0) \gamma_2^A(0) e^{-i\theta_{\sigma}}
\right\}
\\-
R_{\uparrow} R_{\downarrow}
K_2\left[\gamma_1^R(0) \gamma_1^A(0) + \gamma_2^R(0) \gamma_2^A(0)\right]
\\-
\sqrt{R_{\uparrow} R_{\downarrow}D_{\uparrow} D_{\downarrow}}
K_2\left[\gamma_1^R(0) \gamma_2^A(0) + \gamma_2^R(0) \gamma_1^A(0)\right].
\label{B1}
\end{multline}
Here we defined
\begin{equation}
K_i=1- \gamma_i^R(0) \gamma_i^A(0), \quad i=1,2.
\end{equation}
Analogous expressions can also be derived for $A_2$ and $B_2$. Combining Eqs.
\eqref{speq}-\eqref{B1} with the general expression \eqref{current} and
observing that the terms containing the combination $h_a + h_b$ do not
contribute to the current and defining the unity vector in the $x$-direction $\bm{e}_x$, 
we arrive at the final result for the current in the
$S_1$ superconductor
\begin{multline}
\bm{j}(z>0)=\bm{e}_x
\dfrac{e N_0}{2}
\int
d\varepsilon
\left[\tanh \dfrac{\varepsilon}{2T_a} - \tanh \dfrac{\varepsilon}{2T_b}\right]
\\\times
\Biggl<
\theta(v_{x})\theta(v_{z})
v_{x}(R_{\uparrow} - R_{\downarrow})
\\
\left[1-\gamma^R_2(0)\gamma^A_2(0)\right]
\left[1-\gamma^R_1(0)\gamma^A_1(0)\right]
\dfrac{1+\gamma_1^R(z) \gamma_1^A(z)}{1-\gamma_1^R(z)\gamma_1^A(z)}
\\\times
\left[\gamma^R_1(0)\gamma^A_1(0) -\gamma^R_2(0)\gamma^A_2(0) \right]
\Sp\left(\sigma_3 \mathcal{P}\right)
\Biggr>,
\label{js1}
\end{multline}
where
\begin{multline}
\mathcal{P}=
\Bigl|
1-
\left[\gamma^R_1(0)\right]^2 \sqrt{R_{\uparrow} R_{\downarrow}} e^{i\theta_{\sigma}}
-
\left[\gamma^R_2(0)\right]^2 \sqrt{R_{\uparrow} R_{\downarrow}} e^{i\theta_{\sigma}}
\\-
2\gamma^R_2(0) \gamma^R_1(0) \sqrt{D_{\uparrow} D_{\downarrow}} e^{i\theta_{\sigma}}
+
\left[\gamma^R_2(0) \gamma^R_1(0)\right]^2 e^{2i\theta_{\sigma}}
\Bigr|^{-2}.
\label{P}
\end{multline}
The current density in the second superconductor $\bm{j}(z<0)$ is trivially derived from Eq. \eqref{js1} by interchanging the indices $1
\leftrightarrow 2$.

Note that at subgap energies we have $\gamma_i^R(z)\gamma_i^A(z)\equiv 1$ and, hence, the fraction
$[1-\gamma_i^R(0)\gamma_i^A(0)]/[1-\gamma_i^R(z)\gamma_i^A(z)]$ in Eq.
\eqref{js1} becomes indefinite. In this case with the aid of Eqs. \eqref{ric} one can establish an equivalent representation for the above fraction
\begin{multline}
\dfrac{1-\gamma_i^R(0)\gamma_i^A(0)}{1-\gamma_i^R(z)\gamma_i^A(z)}
\\=
\exp\left(
-\dfrac{i\sgn z}{|v_{z}|}
\int_0^z \Delta(z') [\gamma_i^R(z') - \gamma_i^A(z')] dz'
\right),
\end{multline}
which remains regular at subgap energies.

The above equations defining the thermoelectric current density $\bm{j}(z)$ flowing in each of the two superconductors 
$S_1$ and $S_2$ along the spin-active interface represent the main result of this work. It is easy to verify that provided the superconducting 
order parameter in one of the superconductors ($\Delta (z>0)$ or $\Delta (z<0)$) tends to zero, Eq. \eqref{js1} reduces to the result for the 
thermoelectric current in an SN bilayer derived previously \cite{Kalenkov14} within the framework of a different technique.

Let us briefly analyze the above results. To begin with we observe that the thermoelectric current \eqref{js1} may differ from zero only in asymmetric structures consisting of {\it different} superconductors $S_1$ and $S_2$. Furthermore -- similarly to \cite{Kalenkov14} -- the current \eqref{js1}
vanishes if at least one of the two conditions, $D_{\uparrow}=D_{\downarrow}$ or $\theta=0$, is fulfilled. On the other hand, provided both these conditions are simultaneously violated, the thermoelectric current differs from zero and its value can become large.

It is also worth pointing out that the thermoelectric currents \eqref{js1}, \eqref{P} flow in the {\it opposite} directions in the superconductors $S_1$ and $S_2$. In each of the superconductors the current density depends on the coordinate $z$ in the vicinity of the interface and
tends to some nonzero values far from it. The latter feature is specific for our ballistic model within which the elastic mean free path $\ell$ tends to infinity and no electron momentum relaxation occurs. Assuming the mean free path to be finite (which is always the case in any real metal), one can demonstrate that the thermoelectric current density $\bm{j}(z)$ remains appreciable only in the vicinity of the spin-active interface $|z| < \ell$ and decays exponentially
into the superconducting bulk at distances from this interface exceeding the elastic mean free path. The corresponding analysis, however, goes beyond the frames of this work and will be made public elsewhere \cite{KZ14}.

In order to accurately evaluate the general expression for the thermoelectric current defined in Eqs. \eqref{js1}, \eqref{P} one should selfconsistently determine the functions $\gamma^{R,A}_{1,2}(z)$ as well as the order parameter $\Delta (z)$ for any given values of the parameters $D_{\uparrow}$, $D_{\downarrow}$ and $\theta$. Technically this is a rather complicated task which can only be handled numerically. There is, however, no particular need in this calculation since the order-of-magnitude estimate of the current can easily be obtained directly from  Eqs. \eqref{js1}, \eqref{P}. The magnitude of the thermoelectric current density can roughly be estimated as
\begin{equation}
|\bm{j}(z)| \sim j_c (R_{\uparrow} - R_{\downarrow}) \sin\theta \frac{T_a - T_b}{T_c},
\label{jT}
\end{equation}
where $j_c\sim e v_F N_0T_c$ is the critical current of a clean superconductor
with the critical temperature $T_c$. Hence, we observe that, in contrast to the standard situation \cite{Galperin73}, our result (\ref{jT})
does not contain the small factor $T/\varepsilon_F \ll 1$, i.e. the magnitude of the thermoelectric effect becomes really large in our case. For instance, by setting $(R_{\uparrow} - R_{\downarrow}) \sin\theta \sim 1$ and $T_1 - T_2 \sim T_c$, one achieves the thermoelectric
current densities of the same order as the critical one $j_c$.

To conclude, we demonstrated that quasiparticle scattering at spin-active interfaces generates electron-hole imbalance in superconductors which may yield
an enhancement of thermoelectric currents in such structures up to values as high as the critical (depairing) current of a superconductor. The same effect is expected in more complicated (e.g.layered) superconducting structures containing spin-active interfaces. Such thermoelectric currents can easily be detected and investigated in modern experiments.

\end{document}